\documentclass[twocolumn,english]{revtex4-1}
\usepackage[T1]{fontenc}
\usepackage[latin9]{inputenc}
\usepackage{amsmath}
\usepackage{graphicx}
\usepackage{babel}
\begin{document}

\title{Voltage induced by Coriolis force: a new sensing scheme for rotation
velocity}

\author{Yun-He Zhao$^{1}$, Yu-Han Ma$^{2}$}
\email{yhma@csrc.ac.cn}

\selectlanguage{english}%

\address{$^{1}$Department of physics,Beijing Normal University, Beijing 100875,
China~\\
$^{2}$Beijing Computational Science Research Center, Beijing 100193,
China}
\begin{abstract}
We study the motion of the charged particles between a pair of conductor
plates in the non-inertial reference frame. It is found that there
exists a stable voltage between the two conductor plates, which is
proportional to the rotation velocity of the non-inertial system.
This effect is similar to the Hall effect as the result of the Lorenz
force. As an application, we propose a rotation velocity measurement
scheme based on this Coriolis force -induced effect.
\end{abstract}
\maketitle

\section{Introduction}

Rotation velocity sensing is an important part of inertial navigation,
the aim of which is to measure the rotation velocity of non-inertial
system. And it is realized by a instrument named gyroscope \cite{key-1gr,key-9gr2}.
The earliest gyroscope utilizes the precession of the mechanical rotor,
and then the goals of the gyroscope's development are to achieve high
precision and miniaturization. With the development of laser and microelectronics,
people have made optoelectronic gyroscopes, such as Ring Laser Gyroscopes
\cite{key-11fig1,key-16fig2,key-7fig3} and Micro-Electro-Mechanical
System gyroscope \cite{key-9gr2}. In recent years, the development
of nuclear magnetic resonance technology and cold atom technology
leads to continuous improvement in accuracy of the quantum gyroscopes
\cite{key-2qgr1,key-8qgr2,key-3qgr3,key-4qgr4,key-5qgr5,key-qgr6},
the new member of the gyroscope family. Different types of quantum
gyroscopes are also being proposed, such as gyroscope based on nitrogen-vacancy
(NV) color centers \cite{key-1NV} and gyroscope \cite{key-23TFIM}
that utilize the decoherence of the transverse field Ising moedel
(TFIM) in non-initial system.

Interferometric Fiber Optic Gyroscope (IFOG) and Atom Interference
Gyroscope (AIG) \cite{key-15AIG,key-2qgr1}are based on the Sagnac
effect \cite{key-10sagnac,key-13sagnac3}. In 1980 \cite{key-14Lorenz},
Sakurai gave a approach to derive the Sagnac effect by using the similarity
between the Coriolis force and the Lorentz force. With this in mind,
a problem arises here that while the Lorentz force of the charged
particles in magnetic field leads to the Hall effect, then can the
Coriolis force lead to a similar effect? In order to solve this problem,
we study the motion of the charged particles between the conductor
plates in the non-inertial system. It is found that the Coriolis force
would induce a stable voltage between the conductor plates, which
is proportional to the rotation velocity of the system. Thus we use
this effect to design a new kind gyroscope and we call it charging
capacitor gyroscope (CCG). 

This paper is organized as follows. The general motion equation of
charged particles in non-initial reference frame is given in Sec.
II. In Sec. III, we obtain the voltage difference between two conductor
plates when the charged particles moving through them, thus the charging
capacitor gyroscope is put forward . Then, the resolution of CCG with
different structure is discussed in Sec. IV, and conclusions are given
in Sec V.

\section{Motion equation for a charged particle in non-inertial system}

We first consider a charged particle with mass $m$, change $q$ moves
in a non-inertial reference frame with rotation velocity $\overrightarrow{\varOmega}$.
The equation of motion of such a particle reads 

\begin{equation}
m\ddot{\overrightarrow{r}}=q\left(\vec{E}+\overrightarrow{v}\times\overrightarrow{B}\right)+2m\overrightarrow{v}\times\overrightarrow{\varOmega}+m\overrightarrow{\varOmega}\times\left(\overrightarrow{\varOmega}\times\overrightarrow{r}\right)\label{eq:motion}
\end{equation}
where $\overrightarrow{r}$ is the particle's coordinate, $\overrightarrow{E}$
and $\overrightarrow{B}$ are the electromagnetic field. Obviously,
Eq. (\ref{eq:motion}) can be re-written in the Cartesian coordinates
as

\begin{equation}
\begin{cases}
\ddot{x} & =\frac{q}{m}\left(E_{x}+v_{y}B_{z}-v_{z}B_{y}\right)+2\left(v_{y}\varOmega_{z}-v_{z}\varOmega_{y}\right)\\
\ddot{y} & =\frac{q}{m}\left(E_{y}+v_{z}B_{x}-v_{x}B_{z}\right)+2\left(v_{z}\varOmega_{x}-v_{x}\varOmega_{z}\right)\\
\ddot{z} & =\frac{q}{m}\left(E_{z}+v_{x}B_{y}-v_{y}B_{x}\right)+2\left(v_{x}\varOmega_{y}-v_{y}\varOmega_{x}\right)
\end{cases},\label{eq:decompose}
\end{equation}
where the second order terms of $\varOmega$ have been ignored in
Eq. (\ref{eq:decompose}) in the limit $v\gg\varOmega$.

\section{Charged particles moving between conductor plates}
\begin{center}
\begin{figure}
\includegraphics[scale=0.2]{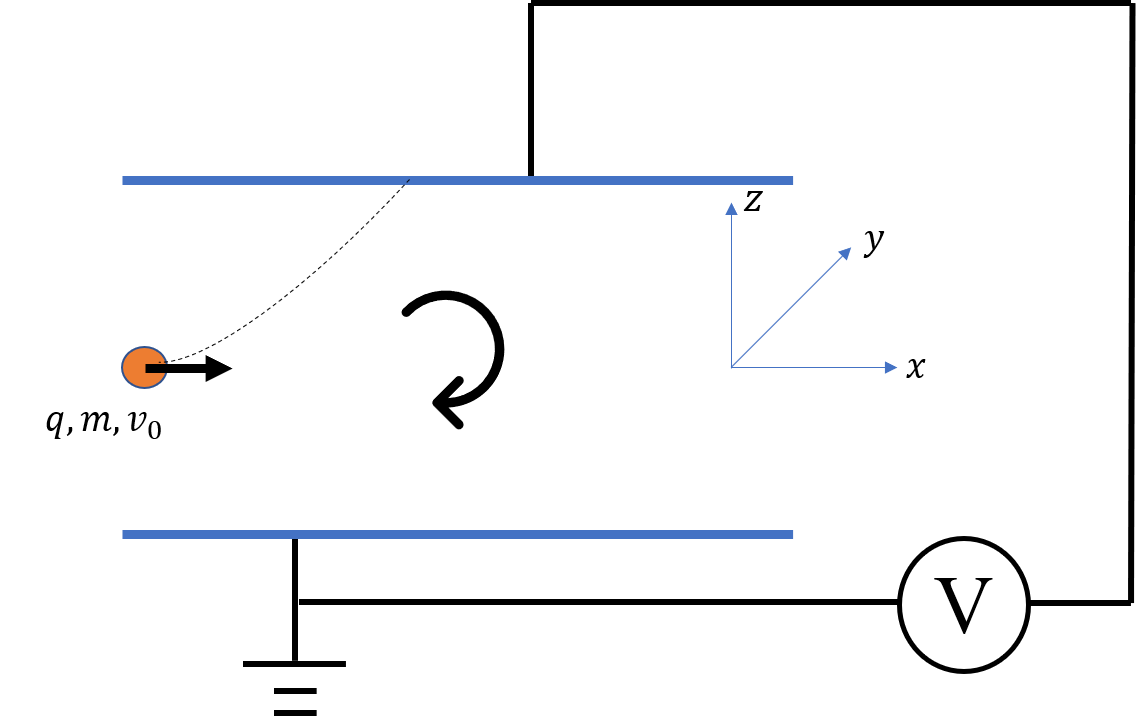}

Figure1. Charged particle flows between conductor plates. The particles
are now moving in $x$ direction with velocity $v_{0}$ between two
conductor plates, where the distance between them is $d$ and The
lower conductor plate is grounded to maintain its potential at zero.
Assuming the system rotates around $y$-direction with rotation velocity
$\varOmega_{0}$, and there is no external electromagnetic field in
the system. 
\end{figure}
\par\end{center}

As shown in Fig. 1, it is obviously that the particles will shift
to the upper plate due to the Coriolis force, thereby charging the
upper conductor plate. In this case, the motion equation for each
particle is

\begin{equation}
\ddot{z}=\frac{q}{m}E_{z}+2v_{0}\varOmega_{0},
\end{equation}
to which the steady solution is

\begin{equation}
E_{z}=\frac{2mv_{0}\varOmega_{0}}{q}.
\end{equation}
For the electric field between the two plates can be approximated
evenly, the stable voltage between the two plates is obtained as

\begin{equation}
U_{z}=E_{z}d=\frac{2mv_{0}\varOmega_{0}d}{q}.\label{eq:voltage}
\end{equation}
This voltage is induced by the Coriolis force of the moving charged
particles in the non-inertial reference frame, and formally similar
to the Hall effect. Once the voltage between the conductor plates
is measured, the rotation velocity of the non-initial system is given
by Eq. (\ref{eq:voltage}) as

\begin{equation}
\varOmega_{0}=\frac{qU_{z}}{2mv_{0}d}.\label{eq:rotation}
\end{equation}
Thus, the measurement of rotation velocity is achieved by this system,
which can be used as a new type of gyroscope, and we call it charging
capacitor gyroscope (CCG).

\section{Resolution of CCG }

It follows from Eq. (\ref{eq:rotation}) that the resolution of CCG
is

\begin{equation}
\triangle\varOmega=\frac{q}{2mv_{0}d}\triangle U,\label{eq:resolution}
\end{equation}
where $\triangle U$ is the resolution of the voltmeter used to detect
the voltage between the two conductor plates. For $\triangle U\sim\mu$V,
$m\sim10^{-27}$kg, $d\sim1$m, $q\sim10^{-19}$C, $v_{0}\sim10^{6}$m/s,
$\triangle\varOmega\sim10^{-4}$rad/s. In order to improve the resolution
of CCG, we put forward an arrangement structure which is shown in
Fig. 2.
\begin{center}
\begin{figure}
\includegraphics[scale=0.2]{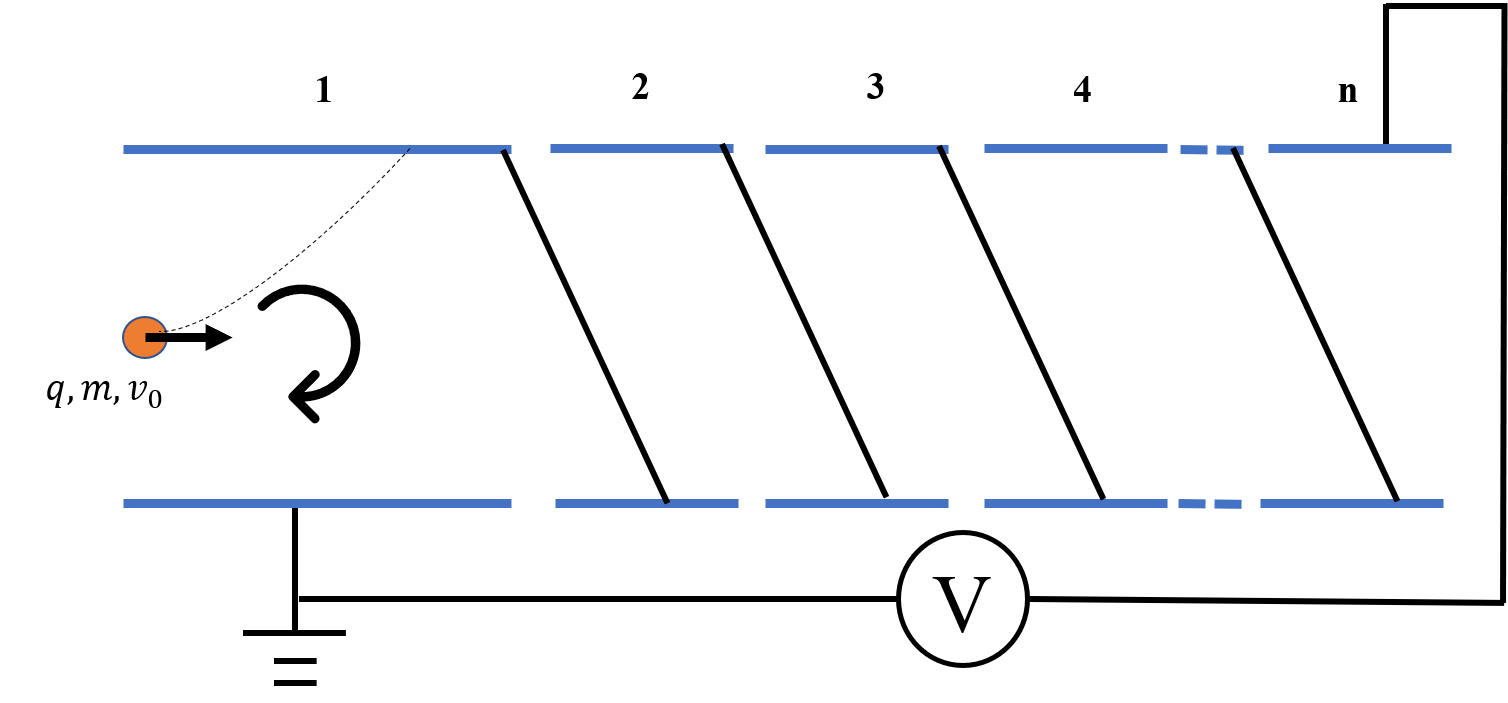}

Figure. 2. Charging capacitor gyroscope (CCG) with Linear structure.
The upper plate of the $i$th conductor plates pair is connect to
the lower plate of the $i+1$th pair with wire to make them have the
same potential.
\end{figure}
\par\end{center}

As shown in Fig. 2, we have 

\begin{equation}
U_{i}^{u}=U_{i+1}^{l}.\label{eq:ii+1}
\end{equation}
Here, $U_{i}^{u}$ ($U_{i}^{l}$ ) is the potential of the upper (lower)
plate belonging to the $i$th conductor plates pair. On the other
hand, for the the $i$th pair , the difference in potential of the
two conductor plates is the same as that in Sec. III, thus

\begin{equation}
U_{i}^{u}-U_{i}^{l}=\frac{2mv_{0}\varOmega_{0}d}{q}.\label{eq:difference}
\end{equation}

Combining Eqs. (\ref{eq:ii+1}) and (\ref{eq:difference}), we have

\begin{equation}
U_{n}^{u}-U_{1}^{l}=\frac{2nmv_{0}\varOmega_{0}d}{q},
\end{equation}
where $n$ is the number of the conductor plates pairs. If we measure
the potential difference between the lower plate of the first pair
of conductor plates and the upper plate of the $n$ th pair, the resolution
of $\varOmega_{0}$ will be

\begin{equation}
\triangle\varOmega=\frac{q}{2nmv_{0}d}\triangle U,
\end{equation}
which is decreased by a factor $1/n$ compared with the result in
Eq. (\ref{eq:resolution}). This indicates that the resolution of
CCG with this structure is $n$ times better than that of the original
one. In the actual production, we can follow the design of cyclotron,
then the above linear cascade structure will changes to a spiral structure,
which can be seen in Fig. 3. 
\begin{figure}
\includegraphics[scale=0.25]{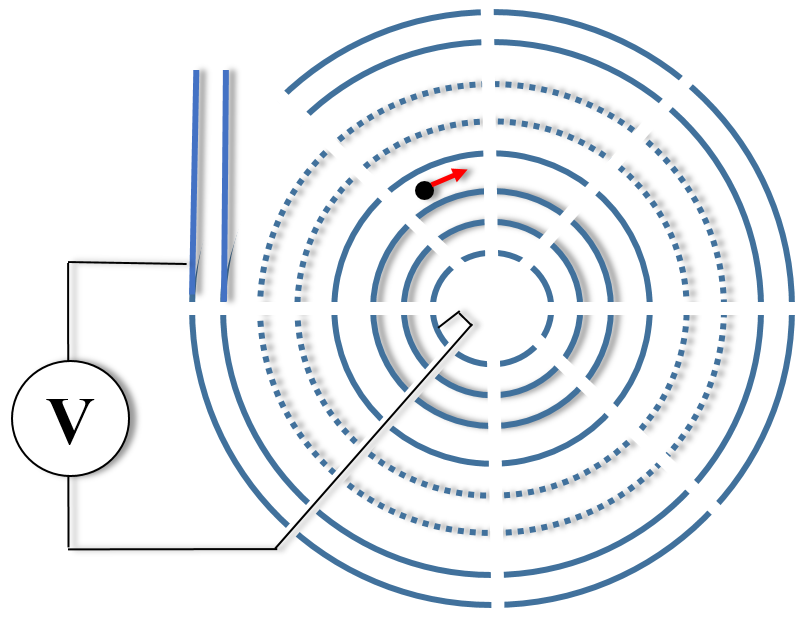}

Figure. 3. Carrier changing gyroscope with helical structure.The distance
and length of each conductor plates pair are $d$ and $l$, and the
radius of the periphery of the disc structure is $R$.
\end{figure}

When $l\ll R$, $d\ll R$, the disc can be loaded with N-layer conductor
plates pairs, thus $N=R/d$. So the total length of conductor is

\begin{equation}
L=\sum_{i=1}^{i=R/d}2\pi id=2\pi d\frac{\left(1+R/d\right)R/d}{2}=\pi R+\frac{S}{d}\approx\frac{S}{d},
\end{equation}
where $S$ is the area of the disc. So the number of conductor plates
pair in this disc structure is $n=L/l=S/dl$. Thus, the expression
of resolution for CCG with the disc structure in Fig. 3 is

\begin{equation}
\triangle\varOmega=\frac{ql}{2Smv_{0}}\triangle U.
\end{equation}
For $S\sim m^{2}$,$l\sim\mu m$, $\triangle\varOmega\sim10^{-10}$rad/s,
which has theoretically reached the resolution of ultra-high-precision
{[}1{]} gyroscopes. However, the structure discussed above also basically
belongs to a two-dimensional structure. So it is possible to further
optimize. By stacking the above-mentioned disc structures in layers,
a three-dimensional structure can be formed. For conductor plates
with height $h$, there are $n^{'}=H/h$ layers of the helical structure
in the three-dimensional structure with height $H$. The connection
conditions for voltage in the three-dimensional structure CCG is thus

\begin{equation}
\begin{array}{c}
U_{i,j}^{u}=U_{i+1,j}^{l}\\
U_{n,j}^{u}=U_{i,j+1}^{l}
\end{array},\label{eq:3-dimention}
\end{equation}
where $i$ and $j$ denote the order of the conductor plates pairs
in each layer and the order of the layers of the spiral structure.
It follows from Eqs. (\ref{eq:difference}) and (\ref{eq:3-dimention})
that

\begin{equation}
U_{n,n^{'}}^{u}-U_{11}^{l}=\frac{2nn^{'}mv_{0}\varOmega_{0}d}{q}=\frac{2HSmv_{0}\varOmega_{0}}{qlh}.
\end{equation}
Now we obtain the resolution of the CCG with the helical columnar
structure as

\begin{equation}
\triangle\varOmega=\frac{qs}{2Vp}\triangle U.
\end{equation}
Here, $s=hl$ is the surface area of each conductor plate, $V=HS$
is the The volume of the entire CCG structure, and $p=mv_{0}$ is
the momentum of each changed particle. Then, for $l,h\sim\mu m$,
$s\sim10^{-12}m^{2}$, $V\sim m^{3}$,$\triangle\varOmega\sim10^{-16}rad/s\sim10^{-11}\textdegree/h$.

\section{Conclusion}

In summary, we found that the charged particles moving between the
conductor plates in the non-inertial system can induce a voltage between
the plates, by measuring which we can obtain the rotational velocity
of the system. This effect gives a new design of microelectronics
gyroscope, which is named as charging capacitor gyroscope (CCG). Inspired
by the analogy of the Coriolis force and the Lorentz force, we have
proposed a sensing scheme for rotational velocity. This protocol is
similar to the magnetic field measurement based on the Hall effect.
By optimizing the structure of the CCG, the best accuracy we have
achieved in this paper is $\triangle\varOmega\sim10^{-11}\textdegree/h$.

\end{document}